\documentclass[12pt]{article}
\usepackage{a4wide,latexsym,graphicx,epsfig,psfrag}
\usepackage{amsmath}
\usepackage{longtable}
\usepackage{amssymb}
\usepackage{cite}

\newcommand{\nn}{\nonumber \\}

\newcommand{\ve}{\varepsilon}

\begin{document}
\parskip=3pt plus 1pt

\begin{titlepage}
\vskip 1cm
\begin{flushright}
{\small\sf  LA-UR 09-03948 \\UWThPh-2009-5 \\
  IFIC/09-24\\FTUV/09-0709\\July 2009} 
\end{flushright}

\vspace*{1.5cm}
\begin{center}
{\Large\bf Reanalysis of pion pion phase shifts \\[.2cm]
from $K \to \pi\pi$  decays} 
\\[20mm]

{\normalsize\bf V. Cirigliano$^{1}$, G. Ecker${^2}$ and
A. Pich$^{3}$}\\
\end{center}

\vspace{1cm}
\begin{flushleft}
${}^{1)}$ Theoretical Division, Los Alamos National Laboratory, Los
 Alamos, NM 87545, USA \\[10pt]
${}^{2)}$ Faculty of Physics, University of Vienna, Boltzmanngasse 5,
A-1090 Wien, Austria \\[10pt]
${}^{3)}$ Departament de F\'{\i}sica Te\`orica, IFIC,
Universitat de Val\`encia -- CSIC, \\
\mbox{} \hspace*{.1cm}  Apt. Correus 22085, E-46071
Val\`encia, Spain
\end{flushleft}

\vfill

\begin{abstract}
\noindent
We re-investigate the impact of isospin violation for extracting the
s-wave $\pi\pi$ scattering phase shift difference $\delta_0(M_K) -
\delta_2(M_K)$ from $K \to \pi\pi$
decays. Compared to our previous analysis in 2003, more precise
experimental data and improved knowledge of low-energy constants are
used. In addition, we employ a more robust data-driven method to
obtain the phase shift difference $\delta_0(M_K) - \delta_2(M_K) =
(52.5 \pm 0.8_{\rm \,exp} \pm 2.8_{\rm \,theor})^\circ $.

\end{abstract}

%
%

\end{titlepage}


\newpage
\addtocounter{page}{1}

\paragraph{1.}
If isospin were conserved the Fermi-Watson final-state interaction
theorem would allow to extract the s-wave pion pion phase shift
difference $\delta_0(M_K) -\delta_2(M_K)$ directly from $K \to \pi\pi$
decay rates. However, the $K \to \pi\pi$
amplitudes are sensitive to isospin violation including
electromagnetic corrections, especially the $I=2$ amplitude
$A_2$. This is due to the large ratio $A_0/A_2 \sim 22$  (octet
enhancement in nonleptonic weak decays) that enhances
isospin-violating corrections to $A_2$.

It has been a long-standing problem to reconcile the phase shift
difference extracted from $K \to \pi\pi$ decays with other
determinations of pion pion phase shifts. This problem has become
especially acute after the precise determination of $\pi\pi$ phase
shifts from combining dispersion theory with chiral
perturbation theory \cite{CGL01}. Our previous analysis of isospin
violation in $K \to \pi\pi$ decays \cite{CENP04} (similar results were
obtained in Ref.~\cite{Bijnens:2004ai}) led to
$\delta_0(M_K) - \delta_2(M_K) = (60.8 \pm 2.2_{\rm \,exp} \pm
3.1_{\rm \,theor})^\circ$, substantially
bigger than the dispersion theoretical result \cite{CGL01}
$\delta_0(M_K) - \delta_2(M_K) = (47.7 \pm 1.5)^\circ$.

We have decided to reanalyse the problem for several reasons.

\begin{itemize}
\item The experimental situation has substantially improved since 2003
 for both the $K^+$ and $K_S$ lifetimes and for the branching ratios
 of $K \to \pi\pi$ decays
 \cite{Ambrosino:2006sh,Antonelli:2008jg,PDG08}. It was already
 observed in Ref.~\cite{flavianetnote} that the new experimental
 information reduces the phase shift difference by more than three
 degrees (with higher statistical significance), bringing it
 closer to the dispersion theoretical value.
\item New information has also become available on some of the
 low-energy constants (LECs) involved, both in the strong
 \cite{Cirigliano:2006hb}  and in the
 electromagnetic sector \cite{Ananthanarayan:2004qk}. The effect on
 the phase shift difference is difficult to quantify but it is
 definitely smaller than the uncertainty assigned to the LECs in
 Ref.~\cite{CENP04}. As one example, the new estimates of
 electromagnetic LECs \cite{Ananthanarayan:2004qk} lead to a shift of
 the lowest-order $\pi^0-\eta$ mixing parameter $\ve^{(2)}$ from
 $1.06\cdot 10^{-2}$ to $1.29\cdot 10^{-2}$
 \cite{Kastner:2008ch}. This shift reduces the phase shift difference
 by $0.2^\circ$.
\item As will be detailed below, the theoretical analysis of
 Ref.~\cite{CENP04} can also be improved by relying to a lesser
 extent on the NLO
 calculation of $K \to \pi\pi$ decay amplitudes. Using less
 information potentially increases the uncertainty but this will be
 compensated by a less biased comparison with the data. The resulting
 estimate of isospin violation is expected to be more robust than the
 original estimate and it leads to a further decrease of the phase
 shift difference by nearly two degrees.
\item Based on a method proposed in Ref.~\cite{Cirigliano:2000zw} making
 use of the optical theorem, effects of $O(e^2 p^4)$ were partially
 accounted for in Ref.~\cite{CENP04}. Since a complete calculation of
 such effects is beyond present capabilities, we have decided to
 keep track of the associated theoretical uncertainty, but without
 including the partial corrections in the mean value for the phase
 shift difference.
\end{itemize}

\paragraph{2.}
We start by recalling the procedure of Ref.~\cite{CENP04}
for extracting the phase shift difference. The amplitudes $A_{+-} =
A(K^0 \to \pi^+ \pi^-)$, $A_{00} = A(K^0 \to \pi^0 \pi^0)$ and
$A_{+0} = A(K^+ \to \pi^+ \pi^0)$ are parametrized as
\begin{eqnarray}
A_{+-} &=&
A_{0} \, e^{i \chi_0}  + { 1 \over \sqrt{2}} \,   A_{2}\,  e^{i\chi_2 }
\nn
A_{00} &=&
A_{0} \, e^{i \chi_0}  - \sqrt{2} \,   A_{2}\,  e^{i\chi_2 }
\label{eq:param1}
\\
A_{+0} &=& {3 \over 2} \,
A_{2}^{+} \,  e^{i\chi_2^{+}}  ~.
\nonumber
\end{eqnarray}
In the absence of CP violation, the amplitudes $A_0, A_2, A_{2}^{+}$
are real and positive by definition. In the isospin limit, $A_2 =
A_{2}^{+}$ and the phases $\chi_I$ coincide with the strong $\pi\pi$
phase shifts $\delta_I$ at the kaon mass.

To NLO in the chiral expansion, the phases $\chi_0, \chi_2$ cannot be
calculated reliably: the resulting phases are substantially too small.
To obtain the strong phase shift difference $\delta_0(M_K) -
\delta_2(M_K)$, a two-step procedure was employed
\cite{CENP04}. Using the NLO expressions for the
absolute values $A_0, A_2, A_{2}^{+}$, the lowest-order couplings
$g_8, g_{27}$ were determined from the experimental rates together
with the phase shift difference $\chi_0 - \chi_2$. Comparing the NLO
amplitudes with and without including isospin violation, the
differences
\begin{equation}
\gamma_I  = \chi_{I} \, - \, \delta_{I}(M_K)  \qquad (I=0,2)
\end{equation}
were calculated to obtain the final phase shift difference
$\delta_0(M_K) - \delta_2(M_K)$ in the isospin limit.

There are two related potential pitfalls associated with this
procedure. Although the theoretical NLO expressions for the phases
$\chi_I$ cannot be trusted they enter the dispersive and absorptive
parts of the amplitudes $A_0, A_2, A_{2}^{+}$ implicitly. Moreover,
although both the
$\chi_I$ and the $\delta_I$ cannot be calculated reliably to NLO in the
chiral expansion the differences $\gamma_I$ were assumed to be
trustworthy. Both
steps are therefore subject to a theoretical bias that is difficult to
control at the level of accuracy considered.

The main idea of the alternative procedure proposed here is to use only
the isospin-violating parts of the NLO amplitudes as theory input and
to determine $\delta_0(M_K) - \delta_2(M_K)$ directly from the data. In
contrast to the chiral corrections for the full amplitudes, the
isospin-violating corrections are much smaller and therefore less
subject to the bias discussed in the previous paragraph.

The amplitudes are now parametrized as
\begin{eqnarray}
A_{+-} &=&
\overline{A}_{0} \, e^{i \delta_0(M_K)}  + { 1 \over \sqrt{2}} \,
\overline{A}_{2}\,  e^{i\delta_2(M_K) } + \Delta A_{+-}^{\rm IB}
\nn
A_{00} &=&
\overline{A}_{0} \, e^{i \delta_0(M_K)}  - \sqrt{2} \,   \overline{A}_{2}\,
e^{i\delta_2(M_K) } + \Delta A_{00}^{\rm IB}
\label{eq:amps}
\\
A_{+0} &=& {3 \over 2} \,
\overline{A}_{2} \,  e^{i\delta_2(M_K)} + \Delta A_{+0}^{\rm IB}  ~.
\nonumber
\end{eqnarray}
All isospin violation is contained in the amplitudes $\Delta
A_{+-}^{\rm IB}, \Delta A_{00}^{\rm IB}, \Delta A_{+0}^{\rm IB}$. They
can be extracted from the NLO amplitudes of Ref.~\cite{CENP04}. Since
isospin violation was neglected in the 27-plet amplitudes because of
the $\Delta I=1/2$ rule \cite{CENP04,Bijnens:2004ai} the
amplitudes $\Delta A_{n}^{\rm IB}$ ($n=+-,00,+0$) scale linearly with
the lowest-order octet coupling $g_8$. They also depend on
higher-order LECs in addition to the loop contributions.

The moduli of the amplitudes in the isospin limit\footnote{We define
the isospin limit in terms of the neutral meson masses \cite{CENP04}.} 
are denoted as $\overline{A}_{0}, \overline{A}_{2}$. We will not use
the theoretical expressions for these amplitudes but instead determine
them together with the phase shift difference $\delta_0(M_K) -
\delta_2(M_K)$ directly from the rates. For this purpose, we write the
moduli of the amplitudes (\ref{eq:amps}) as
\begin{eqnarray}
|A_{+-}| &=&
\left| \overline{A}_{0}  + { 1 \over \sqrt{2}} \,
\overline{A}_{2}\,  e^{i(\delta_2(M_K)- \delta_0(M_K))} +
\Delta A_{+-}^{\rm IB}\,  e^{- i \delta_0(M_K)}\right|
\nn
|A_{00}| &=&
\left| \overline{A}_{0}   - \sqrt{2} \, \overline{A}_{2}\,
e^{i(\delta_2(M_K)- \delta_0(M_K))} + \Delta A_{00}^{\rm IB} \,  e^{- i
 \delta_0(M_K)}\right|
\label{eq:moduli}
\\
|A_{+0}| &=& \left| {3 \over 2} \,
\overline{A}_{2}\, e^{i(\delta_2(M_K)- \delta_0(M_K))}  + \Delta
A_{+0}^{\rm IB} \, e^{- i \delta_0(M_K)}\right|   ~.
\nonumber
\end{eqnarray}
In order to determine $\overline{A}_{0}, \overline{A}_{2}$ and
$\delta_0(M_K)- \delta_2(M_K)$ from the three rates, we therefore also
need the $I=0$ phase $\delta_0(M_K)$ as input. We use the value
obtained in Ref.~\cite{CGL01}:
\begin{equation}
\delta_0(M_K) = (39.2 \pm 1.5)^\circ ~.
\label{eq:d0}
\end{equation}
From the structure of the moduli (\ref{eq:moduli}) one may already
anticipate that the precise value of $\delta_0(M_K)$ will have little
impact on the phase shift difference.

\paragraph{3.}
We use the same experimental input as in Ref.~\cite{flavianetnote}, which
is reproduced in Table \ref{tab:data}.
\begin{table}[here]
\begin{center}
\begin{tabular}{lccc}
\hline\hline
Parameter & Value & Correlation & Reference \\
\hline
$\begin{array}{l}
{\rm BR}(K_S\to\pi^+\pi^-(\gamma)) \\
{\rm BR}(K_S\to\pi^0\pi^0)
\end{array}$ &
$\begin{array}{l}
\mbox{0.69196(51)} \\
\mbox{0.30687(51)}
\end{array}$ &
$-\mbox{0.9996}$ &
\cite{Ambrosino:2006sh} \\
\hline
$\begin{array}{l}\tau_S\end{array}$ &
$\begin{array}{c}\mbox{0.08958(5)~ns}\end{array}$ & & \cite{PDG08} \\
\hline
$\begin{array}{l}
{\rm BR}(K^+\to\pi^+\pi^0) \\
\tau_+
\end{array}$ &
$\begin{array}{c}
\mbox{0.2064(8)} \\
\mbox{12.384(19)~ns}
\end{array}$ &
$-\mbox{0.032}$ &
\cite{Antonelli:2008jg} \\
\hline\hline
\end{tabular}
\end{center}
\caption{Experimental input taken from
 Ref.~\cite{flavianetnote}.}
\label{tab:data}
\end{table}

In addition to the experimental input, we need the isospin-violating
amplitudes $\Delta A_{n}^{\rm IB}$. With the central
values of the various LECs and displaying explicitly the linear
dependence on the octet coupling $g_8$, we find
\begin{eqnarray}
\Delta A_{+-}^{\rm IB} &=& g_8 \, \left[2.25 - 0.83\,i \right] \cdot
 10^{-10} ~~{\rm  GeV} \nn
\Delta A_{00}^{\rm IB} &=& g_8 \, \left[- 0.58 - 2.77\,i \right] \cdot
 10^{-10} ~~{\rm  GeV} \label{eq:IBamps} \\
\Delta A_{+0}^{\rm IB} &=& g_8 \, \left[- 2.12 - 1.10\,i \right] \cdot
 10^{-10} ~~{\rm  GeV} ~. \nonumber
\end{eqnarray}
Updating the fit in Ref.~\cite{CENP04} with the new data, we get a
mean value $g_8 = 3.6$ (see also Ref.~\cite{flavianetnote}). We are
going to assign a 20\% uncertainty to $g_8$, much bigger than the fit
error.

Using the isospin-violating amplitudes (\ref{eq:IBamps}) with
$g_8=3.6$ and $\delta_0(M_K) = 39.2^\circ$ \cite{CGL01}, the
experimental rates in Table \ref{tab:data} give rise to
\begin{eqnarray}
\overline{A}_{0} &=& 2.7030(8) \cdot 10^{-7} ~~{\rm GeV} \nn
\overline{A}_{2} &=& 0.1249(3) \cdot 10^{-7} ~~{\rm GeV} \label{eq:fit}
\\
\delta_0(M_K)- \delta_2(M_K) &=& (52.54 \pm 0.83)^\circ ~. \nonumber
\end{eqnarray}
The errors are purely experimental but they take the correlations in
Table \ref{tab:data} into account. The resulting correlations for the
fitted quantities are small, at most $- 16\% $ for the correlation
between $\overline{A}_{2}$ and $\delta_0(M_K)- \delta_2(M_K)$.

The octet enhancement in $K \to \pi\pi$ decays is characterized by the
amplitude ratio
\begin{equation}
\displaystyle\frac{\overline{A}_{0}}{\overline{A}_{2}} = 21.63(4)~,
\label{eq:ratio}
\end{equation}
again with experimental error only.

\paragraph{4.}
For assessing the theoretical uncertainties, we concentrate on the
phase shift difference.

The uncertainty induced by the error of $\delta_0(M_K)$ in
(\ref{eq:d0}) can be disposed of quickly. Varying the $I=0$ phase
shift as $\delta_0(M_K) = (39.2 \pm 3.0)^\circ ~(2 \,\sigma)$
affects the phase shift difference only in the second decimal place
for $\delta_0(M_K)- \delta_2(M_K)$. Of course, this
has to do with the smallness of the isospin-violating amplitudes
(\ref{eq:IBamps}). The uncertainty in the phase shift difference is
therefore completely negligible.

The overall scale of the isospin-violating amplitudes
(\ref{eq:IBamps}) is determined by the lowest-order coupling $g_8$. We
assign a generous error of 20\%, i.e., $g_8 = 3.6 \pm 0.8$. As
already emphasized, this uncertainty is much larger than the fit error
\cite{CENP04,flavianetnote} but it is meant to account also at least
partially for effects of $O(e^2 p^4)$ (see below). Varying $g_8$ by
20\% gives rise to an uncertainty $\pm 1.1^\circ$ for the phase
difference.

As in Ref.~\cite{CENP04}, we estimate the uncertainty associated with
the various LECs  by varying both
the short-distance scale for the Wilson coefficients (0.77 GeV $\le
\mu_{\rm SD} \le$ 1.3 GeV) and the chiral renormalization scale  (0.6
GeV $\le \nu_{\chi} \le$ 1 GeV). The central values in
(\ref{eq:fit}) correspond to $\mu_{\rm SD} =$ 1 GeV, $\nu_\chi =$ 0.77
GeV. As already mentioned, the changes in the
LECs from 2003 till today are well within the range expected from varying
the two scales. It turns out that the error associated with the
short-distance scale is asymmetrical: the phase difference
happens to be minimal for  $\mu_{\rm SD} =$ 1 GeV and the increase
from varying $\mu_{\rm SD}$ is at most $0.5^\circ$. From the dependence
on the chiral renormalization scale, the error for the
phase difference is bigger and nearly symmetrical around the central
value for $\nu_\chi =$ 0.77 GeV: the phase difference varies by $\pm
1.2^\circ$.

Finally, we consider higher-order effects of $O(e^2 p^4)$.
In Ref.~\cite{CENP04}  we made use of the optical theorem and
$\pi \pi$ scattering amplitudes to  $O(e^2 p^2)$
\cite{Knecht:1997jw,Knecht:2002gz} to  estimate the absorptive
part of the $K \to \pi \pi$ amplitudes  to $O(e^2 p^4)$. Based on that
we quantified the $O(e^2 p^4)$ corrections to
$\gamma_{2}$ (and to $\delta_0(M_K) - \delta_2 (M_K)$)  to be $+ 2.6^\circ$
and included them in our final estimate.
However, since a complete calculation of $K \to \pi \pi$ to
$O(e^2 p^4)$ is not available,
here we assume a more prudent attitude of not including
this correction in the central value. We rather take this result  as a
first measure of the size of higher-order corrections.
We have also reached a similar conclusion on the size of these effects  with no
reference to the optical theorem analysis.
Within the method introduced in this letter,
we have parametrized  higher-order corrections
via three different scale factors  multiplying  the
isospin-breaking amplitudes $\Delta A^{\rm IB}_{+-,00,+0}$.
We have repeated  the fit for different choices of the scale factors ranging
independently  between 0.5 and  1.5,
finding  that the output  $\delta_0(M_K) - \delta_2 (M_K)$ changes by
at most $2.2^\circ$,
which will be used as our final estimate of higher-order effects.

Altogether, our new procedure for confronting theory with experiment
leads to the following final value for the phase shift difference in
the isospin limit:
\begin{eqnarray}
\delta_0(M_K)- \delta_2(M_K) &=&  (52.5 \pm  0.8_{\rm \,exp} \pm
1.1_{\,g_8} ~\mbox{}_{\,- 0.0}^{\,+ 0.5} \,\mbox{}_{\rm SD}  \pm
1.2_{\chi} \pm 2.2_{\, O(e^2 p^4)})^\circ 
\nonumber\\[.2cm]
& = & (52.5 \pm  0.8_{\rm \,exp} \pm  2.8_{\rm \,theor})^\circ ~.
\label{eq:result}
\end{eqnarray}

\paragraph{5.} 

If isospin violation is neglected (except in the physical
pseudoscalar masses for phase space), i.e. taking $\Delta A_{n}^{\rm
  IB}=0$, the fit to the experimental rates gives 
\begin{equation}
\left[ \delta_0(M_K)- \delta_2(M_K)\right]_{\rm Isospin} =  (47.3 \pm
1.0)^\circ  
\label{eq:d02_isospin}
\end{equation}
and
\begin{equation}
\left[
  \displaystyle\frac{\overline{A}_{0}}{\overline{A}_{2}}\right]_{\rm
  Isospin} = 22.41(5)~. 
\label{eq:ratio_isospin}
\end{equation}
The substantial experimental improvements achieved with the most
recent kaon data have reduced  
the phase shift difference from the value $(48.6\pm 2.6)^\circ$
obtained in 2003 in the isospin limit \cite{CENP04} 
(and about $58^\circ$ some 30 years ago \cite{PDG08}). The value 
(\ref{eq:d02_isospin}) would be in perfect agreement with
determinations from $\pi\pi$ scattering data:
\begin{equation}
\left[ \delta_0(M_K) - \delta_2(M_K)\right]_{\pi\pi} =\left\{
\begin{array}{cc}
 (47.7 \pm 1.5)^\circ  &  \qquad\mbox{\protect\cite{CGL01}}
 \\[7pt]
 (50.9 \pm 1.2)^\circ  &  \qquad\mbox{\protect\cite{Kaminski:2006qe}}
 \end{array}\right.
\label{eq:d02_pipi}
\end{equation}

However, owing to the large $A_0/A_2$ ratio in $K\to 2 \pi$ decays,
isospin-breaking corrections to the dominant $\Delta I=1/2$ amplitude
generate sizeable contributions to $A_2$ [compare the results
  (\ref{eq:ratio}) and (\ref{eq:ratio_isospin})],  
modifying also the amplitude phases. Our 2003 analysis of isospin
breaking in $K\to 2 \pi$ decays concluded that these effects increase
the phase shift difference significantly. 
Including some $O(e^2 p^4)$ corrections through the optical theorem, we
found the result $\delta_0(M_K) - \delta_2(M_K) =(60.8 \pm 
 2.2_{\rm \,exp} \pm 3.1_{\rm \,theor})^\circ$
\cite{CENP04}. The large difference with the $\pi\pi$ determinations 
(\ref{eq:d02_pipi}) has been a pending puzzle since then.

In this letter we have reanalysed the $K\to 2 \pi$ phase shift
determination, taking advantage of the improved experimental
situation. Moreover, we have modified the theoretical
analysis in order to be less sensitive to theoretical
uncertainties. Since the absorptive contributions that generate the
strong amplitude phases start to appear at the one-loop level, a NLO
theoretical calculation of the amplitudes only provides the phase
shifts at leading order, which are therefore subject to large
uncertainties. To minimize theoretical errors, we have only used as
theory input the calculation of the isospin-breaking contributions
$\Delta A_{n}^{\rm IB}$. In this way, we can determine all other
quantities (the phase shift difference and the amplitudes in the
isospin limit) directly from a fit to the data.  
The residual theoretical uncertainties associated with the
isospin-breaking contributions have been estimated conservatively in
two different ways, as explained in the previous section. Our final
result in Eq.~(\ref{eq:result}) is lower than our 2003 determination.
Both the new data (as already observed in Ref.~\cite{flavianetnote})  
and the different treatment of theory input tend to lower the
resulting value: $3.3^\circ$ from experiment and altogether 
$5^\circ$ from theory. This  
updated determination of the phase shift
difference from $K\to 2 \pi$ decays turns out to be 
in agreement with the $\pi\pi$ results in (\ref{eq:d02_pipi}),
although with a larger uncertainty.

\section*{Acknowledgements}
We are grateful to Gilberto Colangelo for communicating to us the
value of $\delta_0(M_K)$ obtained in Ref.~\cite{CGL01}.
We would also like to thank Hans Bijnens, J\"urg Gasser, Peter
Minkowski, Helmut Neufeld and Jorge Portol\'es for useful comments.  
This work has been supported in part by
the EU Contract MRTN-CT-2006-035482 (FLAVIAnet),
by MICINN, Spain [grants FPA2007-60323 and Consolider-Ingenio 2010
  CSD2007-00042 –CPAN–] 
and by Generalitat Valenciana (PROMETEO/2008/069).

\end{document}